\documentclass[prl,aps]{revtex4}
\def\b{\begin{equation}}
\def\e{\end{equation}}
\def\l{\left}
\def\r{\right}
\def\l{\left}
\def\r{\right}
\def\o{\over}
\usepackage{amsmath}
\begin{document}

\title{Mass of Schwarzschild  Black Holes Is Indeed Zero And Black Hole Candidates Are Quasi Black Holes }

\author{Abhas Mitra}

\email {akmitra@hbni.ac.in}
\affiliation { Homi Bhabha National Institute, Mumbai-400094, India}




\date{\today}
\begin{abstract}
A Schwarzschild Black Hole (BH) is the gravitational field due to a neutral point mass, and it turns out that the gravitational mass of a neutral point mass: $M=0$ (Arnowitt, Deser, Misner, PRL 4,  375, 1960). The same result is also suggested by  Janis,  Newman, and  Winicour (PRL  20,  878, 1968). In 1969, Bel gave an explicit proof that for a Schwarzschild BH, $M=0$ (Bel, JMP 10,  1051, 1969).  The same result follows from the fact the timelike geodesic of a test particle would turn null if it would ever arrive at an event horizon (Mitra, FPL, 2000, 2002). Non-occurrrence of trapped surfaces in continued gravitational collapse too demands $M=0$ for black hole (Mitra, Pramana, 73, 615, 2009).
 Physically, for a point mass at $R=0$, one expects ${\it Ric} \sim M \delta (R=0)$ (Narlikar \&  Padmanabhan, Found.  Phys.,  18, 659, 1988). But the black hole solution is obtained from ${\it Ric} =0$.  Again this is the most direct proof that $M=0$ for a Schwarzschild black hole. Implication of this result is that the observed massive black hole candidates are non-singular quasi black holes or black hole mimickers which can possess strong magnetic fields as has been observed. The echoes from LIGO signals, if true, may be the direct evidence that the pertinent compact objects are BH mimickers and not true vacuum BHs.

\end{abstract}




\maketitle

\section{Introduction}

Starting from Einstein, most of the founding fathers of general relativity thought that though the vacuum Schwarzschild equation is correct, its interpretation in terms of black holes is incorrect. Essentially, they pleaded that while a massive star can contract almost upto its Schwarzschild surface, it cannot contract upto the redline $R=R_s = 2M$ ($G=c=1$), and its radius $R > 2M$. Even when the mathematical BH paradigm got well established by early 1960s, Dirac ruled out the concept of point singularities \cite{1}:

``The mathematicians can go beyond this Schwarzschild radius, and get inside, but I would maintain that this inside region is not physical space, because to send a signal inside and get it out again would take an infinite time, so I feel that the space inside the Schwarzschild radius must belong to a different universe and should not be taken into account in any physical theory. So from the physical point of view, the possibility of having a point singularity in the Einstein field is ruled out. Each particle {\em must have a finite size no smaller than the Schwarzschild radius}."

From physical perspective, contrary to the myth, BH paradigm got never established because the physical nature of BH interiors have never been understood. This becomes clear from the following admission by three physicists who believe in the BH paradigm. These authors commented about the coordinate transformations which endeavour to present the BH interior as regular region of spacetime \cite{2}:

“The solutions that do away with the interior singularity and the event horizon, although interesting in themselves, sweep the inherent conceptual difficulties of black holes under the rug. In concluding, we note that the interior structure of realistic black holes have not been satisfactorily determined, and are still open to considerable debate.”

And the reason that the BH interior is not understood after more than a century of the discovery of the Schwarzschild (actually Droste -Hilbert) solution is that {\em there is no true BH, and no BH interior}. In other words true mathematical Schwarzschild BH has $M=0$, and the so-called BHs are only BH mimickers or quasi-BHs. While we established this result from various independent lines of aruments since 2000, in 2009, we offered a more direct proof \cite{3}. This proof was based on the 170 year old rule of multivariate calculus first obtained by Carl Jacobi (1804 -51 CE):  coordinate transformation from say from $x, y$ to $u(x, y), v (x, y)$, obeys the rule
\b
dx dy = dy dx= J  du dv = J dv du
\e
where

\[
J =
\det{\begin{vmatrix}
x_u & y_v \\
x_v  & y_u \\
\end{vmatrix}}
\]

   It may be  noted that the inherent anti-symmetrization needed for defining area or volume is already incorporated in the definition of $J$. Here $x_u$ denotes partial differentiation of $x$ by $u$ and so on. Note,  $dx, dy, du, dv$ are infinitesimal  increments of the corresponding variables, and $dx dy$ implies ordinary symmetric multiplication.

However now Kundu has claimed that this 170 year old rule is incorrect unless one would  reinterpret this in terms of anti-symmetric product of corresponding one forms \cite{4}:

Eqs.(1), (3) and (4) though discussed in all text books are incorrect! Note  while Eqs. (3) and (4) define only the amount/magnitde of a volume element, subsequent delevelopments in mathematics allowed for an {\em oriented volume form} in terms of wedge products of one-forms which is something like covariant vectors. While wedge product is similar to vector cross products in 2-D, in general it is different. Even in 2-D, wedge product results in 2-headed tensors  (2-forms),  cross-product leads to a single (polar) vector. In the context of GR, 4-D invariant volume form is

\b
{\bf dx} \wedge  {\bf dy} = - {\bf dy} \wedge {\bf dx}  =  J  {\bf du} \wedge {\bf dv} =- J {\bf dv} \wedge {\bf du}
\e

For a moment, let us accept that this is true. If so, the direct proof offered ceases to be valid. From this Kundu has jumped to the conclusion that all other independent proofs which show $M=0$ for a BH too must be incorrect. But we reiterate that there is no basis for such illogical sweeping conclusion.

\subsection{Scharzschild Singularity: Sphere or Point?}
 Earlier "Event Horizon" was known as "Schwarzschild Singularity". And long back  Janis,  Newman, and  Winicour \cite{5} wrote

``A spherically symmetric solution of the Einstein equations is presented that coincides
with the exterior   Schwarzschild solution'' as long as $R> 2M$. But as soon as one would attempt to have a solution for $R= 2M$, one would obtain a {\em point singularity instead of Schwarzschild surface}''. Since this behaviour persists for arbitrary weak coupling of the massless scalar field, they wondered whether there could ever be a Schwarzschild surface with finite radius \cite{5}:

``It is clear that if our truncated Schwarzschild
metric is to be considered as the physical solution
corresponding to a, spherically symmetric
point mass, {\em then the entire question of gravitational
collapse beyond the Schwarzschild radius
becomes meaningless}. This point of view also
obviates all discussion of the topological questions
of the Schwarzschild interior, which for
many people has always been disturbing.''

And immediately afterwards, Bel offered explicit proof that the Schwarzschild surface behaves like a point singularity \cite {6}: 

``A new point of view is presented for which the Schwarzschild singularity becomes a real point singularity
on which the sources of Schwarzschild's exterior solution are localized.''

Bel showed that as if the Scharzschild surface itself is the source of a point mass having a Dirac delta energy momentum tensor. Implication of his result though ignored and not explored further is that the point mass has $M=0$ as found by us.

Even if one would ignore Bel's paper, on physical grounds, the energy momentum tensor of the point mass or black hole singularity should be infinite. An explicit form of such an Dirac delta energy momentum tensor was first given by   Tangherlini \cite{7}. On the other hand, Narlikar \& Padmabhan \cite{8}  considered a simpler form of the energy momentum tensor of the point mass with an attendant

\b
{\it Ric} \sim M \delta (R=0)
\e

However, as is well known, the vacuum Schwarzschild - Hilbert solution is obtained from the condition

\b
{\it Ric} =0
\e
everywhere including at $R=0$. Then compatibility of these two foregoing equations directly demands that $M=0$ for a point mass, so that the Schwarzschild Singularity ($R=2M$) is a point rather than than a sphere.

\section{Is There Any Exact Solution for BH Formation?}
Starting from Einstein, most of the founding fathers of general relativity opined that the Schwarzschild Singularity at $R=2M$ should not occur in  Nature and gravitational collapse should result in compact objects having radius $R > 2M$. But this view apparently got contradicted by the Oppenheimer -Snyder study which is widely cited as evidence in favor of formation of finite mass black holes \cite{9}. However what is almost universally overlooked here is the fact that Oppenheimer \& Snyder thought that black holes or singularities should not form in realistic collapse.  They  admitted that they were compelled to assume pressureless collapse in order to obtain an exact treatment and occurrence of black hole was due to such idealistic and unphysical simplification \cite{9}:

``it Physically such a singularity
would mean that the expression used for the
energy-momentum tensor does not take account
of some essential physical fact which would
really smooth the singularity out.''

So they were keenly aware that such a mathametical singularity would be a result of ``not taking into account some essential physical fact'' like pressure gradient, and flow of heat and radiation. Finally they wrote \cite{9}

`` Further, a star
in its early stage of development would not
possess a singular density or pressure; it is
impossible for a singularity to develop in a finite
time.''

By this, they asserted that {\em for a realistic case, no singularity should form ever}.

 It was indeed shown that Oppenheimer \& Snyder black hole tacitly corresponds to $M=0$ and which never forms as the comoving proper time $\tau =\infty$ in such a case \cite{10, 11}. 
Even when one would use Kruskal coordinates, it has been shown that, Schwarzschild BHs corresponds to $M=0$\cite{12}.

\section{Persistent Objection to Idea of BHs}

Despite the idea that the Schwarzschild surface is only a coordinate singularity, many leading general relativists have contested the validity of the coordinate transformations which purport to earse the metric singularity at $R=2M$ on following arguments \cite{13, 14, 15, 16, 17, 18, 19, 20, 21,22,23}:

$\bullet$ The original problem of a static point mass was by definition a static one. But all such coordinate tranformations render the problem non-static.

$\bullet$ Though it might be mathematically acceptable to declare that, inside the event horizon, `space' and `time' coordinates interchange their role, this is not physically meaningless, but inconsistent too because originally one defines $R$ as a spacelike coordinate. This is also inconsistent because even for $R < 2M$, $g_{\theta \theta}$ and $g_{\phi \phi}$ retain their original signatures. Had `space' would really be `time' like, $\theta$ and $\phi$ too have become time like.

$\bullet$ The metric determinant associated with such singularity erasing coordinates {\em become singular} at $R=2M$ which is not allowed, and as if such exercises hide one infinity with another.

$\bullet$ Inside the Schwrzschild surface, infalling test particles will be superluminal and suddenly behave as Tachyons.

\section{Evidence For Physical Singularity}
 
Under the assumption $M >0$, the Kretschmann scalar is of course finite at $R=2M$: $K_{EH} = (3/4) M^{-4}$. Even when one accepts this argument for declaring event horizon as a mere coordinate singularity, this surface continues to display physically significant singular properties. First singular property of the event horizon is that photons emitted from there suffers infinite gravitational redshift.  This is  a coordinate independent measurable effect and which really makes black holes `black'. In 1989, Padmanabhan \cite{24} showed that during contraction, the phase volume and density of states of  a star would blow up as soon as it would dip down to $R = 2M$ surface. It may be noted that phase volume determines all 
thermodynamical quantities of the star. Padmabhan emphasized that \cite{24}

 ``The divergence is
due to the divergence of locally measured momentum
space volume near $R \approx 2M$ (and {\em not due to any
coordinate space divergence}). It is usual to dismiss
the divergence of  $g_{TT} =\infty$ at $R= 2M$, as a “mere coordinate
effect”; our analysis persuades one to rethink
about such “mere” coordinate effects.''

 Independently,  Borkar and Karade \cite{25} have claimed that Schwarzschild singularity is a physical singularity and not a mere coordinate singularity.

\subsection{Direct Evidence For Singular Nature}

The proper acceleration of a test particle is obtained by taking norm of the acceleration 4-vector and is a scalar \cite{26}:

\b
a = {M\over R^2 \sqrt{1- 2M/R}}
\e

This scalar is measureable by an accelerometer and diverges at $R= 2M$. Also since this measureable scalar would become imaginary for $R < 2M$ one may conclude  that there should not be any black hole interior, i.e., one should have $M=0$.

In 1982, Karlhede et al., investigated the properties of local geometry of Scharzschild space time in terms of the Riemann curvature tensor and its higher derivatives \cite{27}. They discovered that the lowest-order non-trivial scalar term, constructed by contracting the covariant derivative of the curvature tensor with itself, has surprising properties.  Karlhede’s invariant can be written a

\b
{\cal I} = R^{ijkl;m} R_{ijkl;m} = {-720 M^2 (R -2M) \o R^9}
\e

This ``Karlhede Invariant'' is a measurable quantity and  it not only becomes singular at $R= 2M$ but would change sign if there would be any black hole interior. And contrary to the notion behind  a ``coordinate singularity'', a free falling observer can very well detect the event horizon by measuring ${\cal I} =0$ or noting its change of sign. This strongly suggests that the event horizon is a physical singularity in accordance with the result that $M=0$ and $K_{EH} = (3/4) M^{-4} = \infty$.

\section{Independent Evidence That $M=0$ For Theoretical BHs}

 For a material particle test moving radially inward towards a black hole, the coordinate speed is given by \cite{28, 29}

\b
\l({dR\o dT}\r)^2 = {(1- 2M/R)^2 \o E^2} \l[ E^2 - (1 - 2M/R)\r]
\e

As $R \to 2M$, this equation reduces to

\b
\l({dR\o dT}\r)^2 \to (1- 2M/R)^2  
\e

And this is exactly what gives the coordinate speed of a radially moving photon. This suggests that material particle would behave like a photon at $R=2M$. Note Eq. (20) leads to

\b
(1- 2M/R) dT^2 \to (1- 2M/R)^{-1} dR^2; ~ R \to 2M
\e

 Further since, for radial motion of the material particle ($d\theta = d\phi =0$), by feeding  the foregoing equation into the vacuum Schwarzschild equation, we find that \cite{27, 28}

\b
ds^2 = (1 - 2M/R) dT^2 - {dR^2 \o 1 - 2 M/R} \to 0 ~as ~ R \to 2M
\e

This is a confirmation, that the worldline of a material particle would become null if it would hit the event horizon. But since this is not allowed, and worldline of the material particle must remain timelike, one must have $\tau_{EH} = \infty$ \cite{28}.

For a test particle starting its journey from $R= R_b$, the comoving proper time to reach the event horizon is

\b
\tau_{EH} = {2\o 3} {\pi R_b^{3/2} \o \sqrt{2M}} - \pi M
\e 

and which can be infinite only if mass of the black hole $M=0$.

Finally the proof that during gravitational collapse, trapped surfaces or apparent horizon must not form \cite{30, 31, 32}
\b
{2M(r, t)\o R} \le 1
\e

in order that the worldlines of stellar material must remain timelike too shows that true finite mass black hole should never form, and at the most, one can have $M=0$ black hole formed after infinite proper time.

\section{Conclusions}
There is no escape from the result that a neutral point mass has $M=0$ and which corresponds to the mass of the so-called Schwarzschild BHs (actually Droste - Hilbert BHs).   We presented here several direct or indirect evidences that indeed $M=0$ for Schwarzschild- Hilbert black hole. The occurrence of $M=0$ does not imply absence of matter, on the other hand it indicates a state where the bare mass comprising rest masses and internal energies are neutralized by negative gravitational energy.  This may be understood by the Arnowitt, Desar and Misner prescription about the gravitational mass $M$ and bare mass $M_b$ of a tiny static sphere of  radius $\epsilon$ ($c=1, G=G$) \cite{33}:

\b
M = G^{-1} \l( -\epsilon + [\epsilon^2 + 2 M_b G \epsilon]^{1/2} \r)
\e

In particular, this paper suggests that for a point charge, one has $M= 2 |e|$ \cite{33}. And thus for a neutral point having $|e| =0$, one has $M=0$. This however shows that, mathematically, chareged BHs may possess finite gravitational mass whose origin is electrical charge.

Coming back to chargeless cases, for a contracting star, $M >0$ as long as $\epsilon > 2 GM_b$. But in the limit $\epsilon \to 0$, one has $M \to 0$, so that, the gravitational mass of a neutral point particle $M=0$. If so, the density of a dust, a collection of point particles, is zero and which explains why we obtained $M=0$ for the Oppenheimer - Snyder black hole \cite{10, 11, 12}.

A very pertinent question is then what are the true nature of the black hole candidates detected by the astronomers. Well, in 1988, Narlikar and Padmanabhan wrote that \cite{8}

``...the discussion of physical behavior of black holes, classical or quantum, is only of academic interest. It is suggested that problems related to the source could be avoided if the event horizon did not form and that the universe only contained quasi-black holes.''

They however had no idea how continued gravitational collapse may give rise to quasi - black holes instead of true mathematical black holes. In contrast,  long ago, McCrea \cite{13} had a worthwhile insight how this might happen. He considered the possibility that if a body would contract to the limit $R \to 2M$, then it may radiate out even $100 \%$ of the original mass-energy so that the final mass may approach $M \to 0$ and exact $M=0$ not forming ever:

``Thus if we tried to make a body of the Schwarzschild critical radius we could in prin­
ciple get as near to doing so as we wish, but we could never quite finish it. So the problem of 
the properties of a body having exactly the Schwarzschild radius is entirely academic.''

Series of research papers by the present author have shown that McCrea's insights were largely correct. In general, pressure  and density of the collapsing matter increases during gravitational contraction. And increase of pressure and its gradient  actually reduces the active gravitational mass and opposes tendency for formation of trapped surfaces \cite{34, 35}. Also, a  self-gravitating body must radiate during contraction and yet get hotter and hotter \cite{35}.  During this process,  the lumonosity of the gravitationally trapped radiation  must attend its general relativistic Eddington limit when the outward radiation pressure exactly balances the inward pull of gravitation \cite{ 37, 38, 39, 40}. Indeed, {\em there are several semi-qualitative and numerical researches which shown that physical gravitational collapse may result in ultra-hot balls of radiating plasma in lieu of any black hole or naked singularity} \cite{41, 42, 43, 44, 45, 46}.  At this stage, the interior of the body is extremely hot and for stellar mass cases, the mean local temperature could be $\ge 200$ MeV so that the matter is in a state of quark gluon plasma. Thus instead of a static black hole, very massive objects end up as quasi-static balls of ultra-hot plasma. This happens at a surface gravitational redshift $z >>> 1$ and in the absence of an exact horizon, the body continues to radiate and contract at infinitesimal rate in order to asymptotically attain the exact black hole stage with $z=\infty$ and $M=0$. Accordingly, such an object has been termed as `Eternally Collapsing Object' (ECO) \cite{38,  38, 39, 40}.

Since stellar mass ECOs are much more compact and relativistic than, neutron stars, in view of  freezing of magnetic flux, their intrinsic magnetic fields are expected to be much higher than that of neutron stars. Thus, ECOs are predicted to be ultra-magnetized and also called Magnetospheric ECOs or MECOs. Many observed properties of black hole x-ray/radio binaries have been explained in this paradigm \cite{46, 47, 48, 49, 50, 51, 52}. There have been also indirect evidences that quasars contain MECOs instead of black holes which have no intinsic magnetic fields \cite{ 44, 45, 46, 47, 48, 49}.  The  ambient magnetic fields around true accreting astrophysical black holes arising from accretion flows or accretion disks are expected to be much weaker and un-organized than what have been found for so-called astrophysical black holes \cite{50, 51, 52, 53, 54, 55}. And such observations point to the fact that $M=0$ for true black holes and so-called black holes are only quasi black holes. Finally one may argue that the detection of gravitational waves by LIGO have confirmed existence of true black holes. But  it turns out that LIGO signals if genuine correspond to detection of "Photon Spheres"  ($R=3M$) and to event horizons ($R=2M$) \cite{56}. Also, it has been found that LIGO signals are corrupted by unknown noise and the LIGO claims of detection of gravitational waves may not be genuine \cite{57}.

And in case, the LIGO claims of detection of gravitational waves are true, they may be confirming that the pertinent compact objects are BH mimickers and not true BHs. This is so because LIGO signals may be superimposed by echoes too \cite{58}. And there cannot be any echo of gravitational or any other wave from true BHs which are vacuum surrounded by event horizons. And thus potential detection of echoes might be the direct evidence for non-existence of true BHs.


\end{document}